\def \SAIT #1 #2 {{\em Mem.\ Soc.\ Astron.\ It.\/} {\bf #1}, #2}
\def \MESS #1 #2 {{\em The Messenger\/} {\bf #1}, #2}
\def \ASTRNACH #1 #2 {{\em Astron. Nach.\/} {\bf #1}, #2}
\def \AAP #1 #2 {{\em Astron. Astrophys.\/} {\bf #1}, #2}
\def \AAL #1 #2 {{\em Astron. Astrophys. Lett.\/} {\bf #1}, L#2}
\def \AAR #1 #2 {{\em Astron. Astrophys. Rev.\/} {\bf #1}, #2}
\def \AAS #1 #2 {{\em Astron. Astrophys. Suppl. Ser.\/} {\bf #1}, #2}
\def \AJ #1 #2 {{\em Astron. J.\/} {\bf #1}, #2}
\def \ANNREV #1 #2 {{\em Ann. Rev. Astron. Astrophys.\/} {\bf #1}, #2}
\def \APJ #1 #2 {{\em Astrophys. J.\/} {\bf #1}, #2}
\def \APJL #1 #2 {{\em Astrophys. J. Lett.\/} {\bf #1}, L#2}
\def \APJS #1 #2 {{\em Astrophys. J. Suppl.\/} {\bf #1}, #2}
\def \APSS #1 #2 {{\em Astrophys. Space Sci.\/} {\bf #1}, #2}
\def \ASR #1 #2 {{\em Adv. Space Res.\/} {\bf #1}, #2}
\def \BAIC #1 #2 {{\em Bull. Astron. Inst. Czechosl.\/} {\bf #1}, #2}
\def \JSQRT #1 #2 {{\em J. Quant. Spectrosc. Radiat. Transfer\/} {\bf #1}, #2}
\def \MN #1 #2 {{\em Mon. Not. R. Astr. Soc.\/} {\bf #1}, #2}
\def \MEM #1 #2 {{\em Mem. R. Astr. Soc.\/} {\bf #1}, #2}
\def \PLR #1 #2 {{\em Phys. Lett. Rev.\/} {\bf #1}, #2}
\def \PASJ #1 #2 {{\em Publ. Astron. Soc. Japan\/} {\bf #1}, #2}
\def \PASP #1 #2 {{\em Publ. Astr. Soc. Pacific\/} {\bf #1}, #2}
\def \NAT #1 #2 {{\em Nature\/} {\bf #1}, #2}

\documentstyle{memsait}
\input epsf.sty
\begin{opening}
\title{THE DYNAMICAL EVOLUTION OF THE DUST SHELL OF IRC\,+10\,216
} % ALL CAPITAL LETTERS PLEASE !!!
\author{R.\,OSTERBART$^1$,
Y.\,BALEGA$^2$,
T.\,BL\"OCKER$^1$,
A.\,MEN'SHCHIKOV$^{3}$,
G.\,WEIGELT$^1$}
\institute{
$^1$Max--Planck--Institut f\"ur Radioastronomie, Bonn, Germany\\
$^2$Special Astrophysical Observatory, Nizhnij Arkhyz, Russia\\
$^3$Stockholm Observatory, Saltsj\"obaden, Sweden
}
\date{} % DO NOT INSERT ANY DATE HERE !!!
\end{opening}

\begin{document}

%\oddpagefooter{\sf Mem. S.A.It., Vol. ??, ??}{}{\thepage}
%\evenpagefooter{\thepage}{}{\sf Mem. S.A.It., Vol. ??, ??}
\oddpagefooter{}{}{} % LEAVE AS IT IS !
\evenpagefooter{}{}{} % LEAVE AS IT IS !
\ 
\bigskip

\begin{abstract}
We present high--resolution $J$--, $H$--, and $K$--band observations of
the carbon star IRC\,+10\,216.  The images were reconstructed from 6 m
telescope speckle interferograms using the bispectrum speckle
interferometry method.  The $H$ and $K$ images consist of several
compact components within a 0.2" radius and a fainter asymmetric nebula.
The brightest four components are denoted with A to D in the order of
decreasing brightness.
A comparison of our images %%% from 1995, 1996, 1997, and 1998
gives --- almost like a movie of five frames --- insight to the dynamical
evolution of the inner nebula. For instance, the separation of the two
brightest components A and B increased by almost 40$\%$ from 191 mas in
1995 to 265 mas in 1998.  At the same time, component B is fading and
the components C and D become brighter.  The {\sf X}--shaped bipolar
structure of the nebula implies an asymmetric mass--loss suggesting that
IRC\,+10\,216 is very advanced in its AGB evolution, shortly before
turning into a protoplanetary nebula.  The cometary shape of component A
suggests that the core of A is not the central star, but the southern
lobe of a bipolar structure.  The position of the central star is
probably at or near the position of component B.
\end{abstract}

\section{Introduction}
IRC\,+10\,216 (CW Leo) is the nearest and best--studied carbon star and
one of the brightest infrared sources. %%% in the sky.
It experiences 
strong mass-loss rates of $\dot{M}
\approx 2-5\times10^{-5}$M$_{\odot}\,$yr$^{-1}$ (Loup et~al.\
1993\nocite{LoupForveilleEtAl93}).  The central star of IRC\,+10\,216 is a
long--period variable with a period of
$\sim 649$\,days (Le~Bertre 1992\nocite{LeBertre92}).
Recent distance estimates range from 
110~pc to 150~pc (Groenewegen 1997\nocite{Groenewegen97}, Crosas
\& Menten 1997\nocite{CrosasMenten97}).
IRC\,+10\,216's initial
mass can be expected to be close to 4\,M$_\odot$ (Guelin et~al.\
1995\nocite{GuelinForestiniEtAl95}, Weigelt et~al.\
1998\nocite{WeigeltBalegaEtAl98}).  The bipolar appearance of the nebula
around this object was already reported, e.g., by Kastner \&
Weintraub \cite{KastnerWeintraub94}.  The non-spherical structure is
consistent with the conjecture that IRC\,+10\,216 is in a phase
immediately before entering the protoplanetary nebula stage.  The most
recent high--resolution observations of this object and its
circumstellar dust shell were reported by
%%% Osterbart et~al. (\cite{OsterbartBalegaEtAl97}),
Weigelt et~al. \cite{WeigeltBalegaEtAl98},
%%% \cite{WeigeltBloeckerEtAl99}),
Haniff \& Buscher \cite{HaniffBuscher98}, and Tuthill et~al.\
\cite{TuthillMonnierEtAl99}.  The results of Dyck et~al.\
\cite{DyckBensonEtAl91} and Haniff \& Buscher \cite{HaniffBuscher98}
showed
that the dust-shell structure of IRC\,+10\,216 is changing
within some years.

\section{Observations and bispectrum speckle interferometry results}
The IRC\,+10\,216 speckle interferograms were obtained with the 6\,m
telescope at the Special Astrophysical Observatory in Russia.  At all
continuum
epochs data within the $K$--band were obtained (date / center wavelength
of the filter in $\mu$m / FWHM bandwidth of the filter in $\mu$m:
8.10.95/2.12/0.02,
3.4.96/2.17/0.02,
23.1.97/2.19/0.41,
14.6.98/2.17/0.33,
3.11.98/2.19/0.19).
$J$- and $H$-band data were recorded at one epoch (2.4.96/1.24/0.28,
23.1.97/1.64/0.31) each.

\begin{figure}
  \setlength{\unitlength}{0.5mm}
  \begin{picture}(146,365)(0,0)
    \put(0,292){ \epsfxsize=36mm\epsfbox{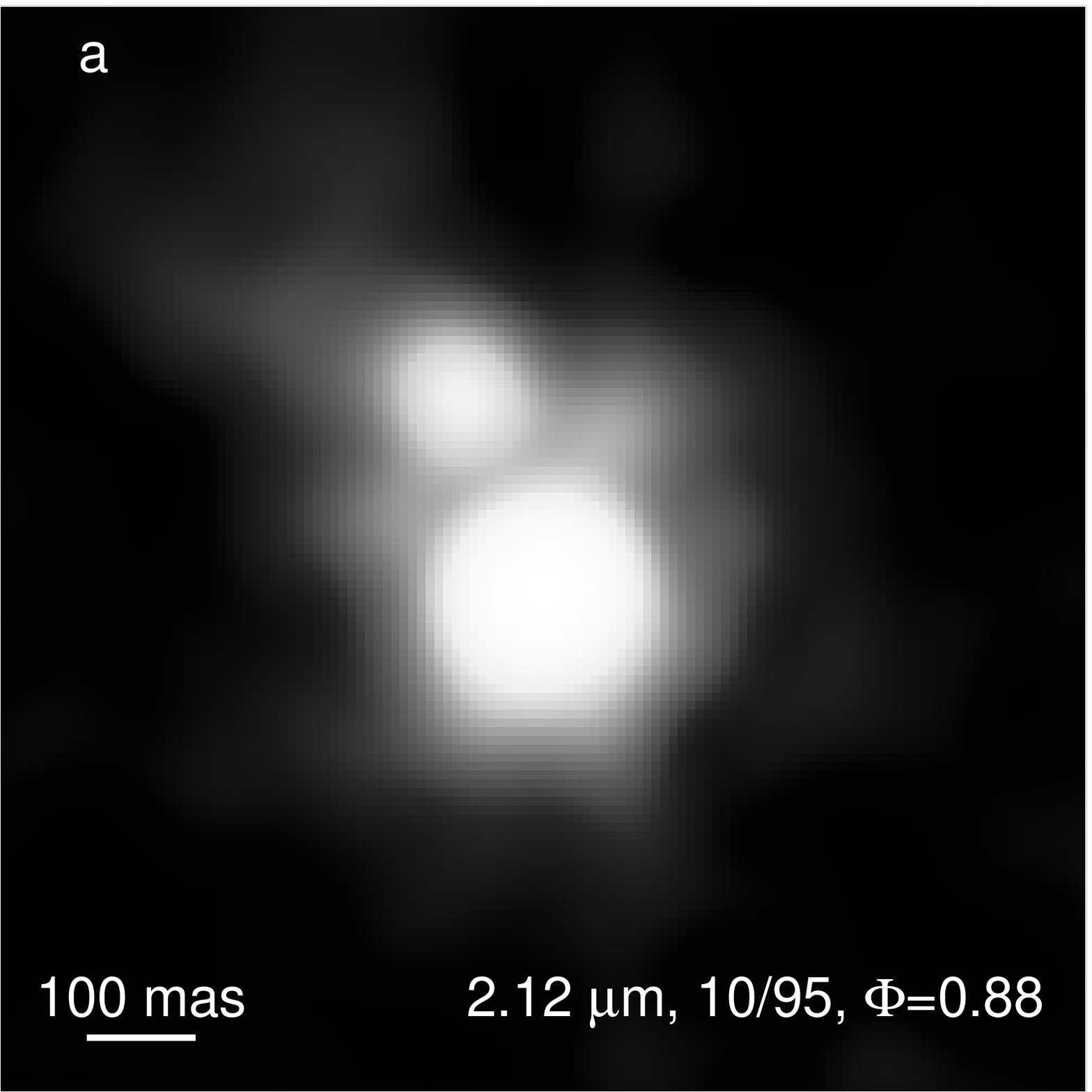}    }
    \put(0,219){ \epsfxsize=36mm\epsfbox{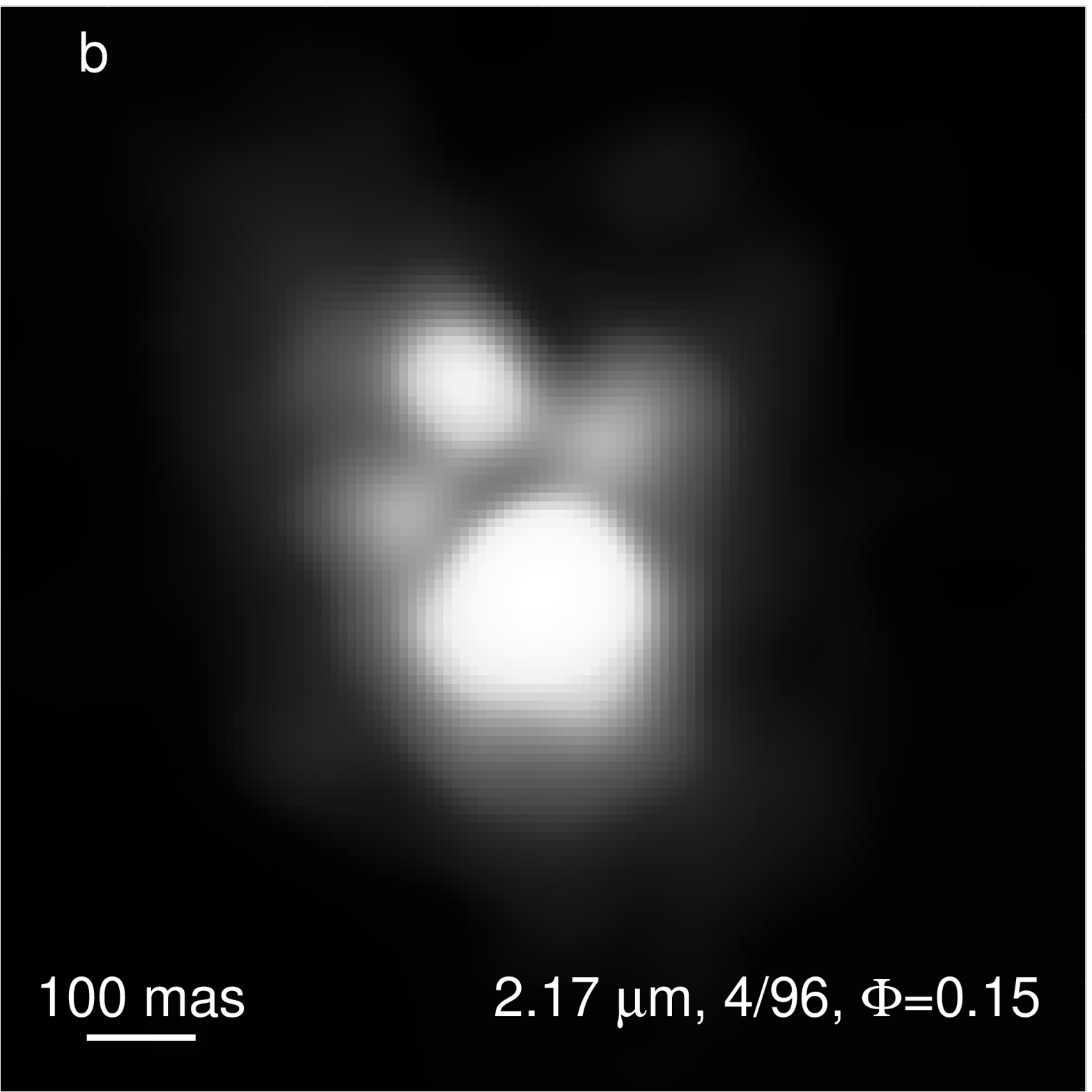}    }
    \put(0,146){ \epsfxsize=36mm\epsfbox{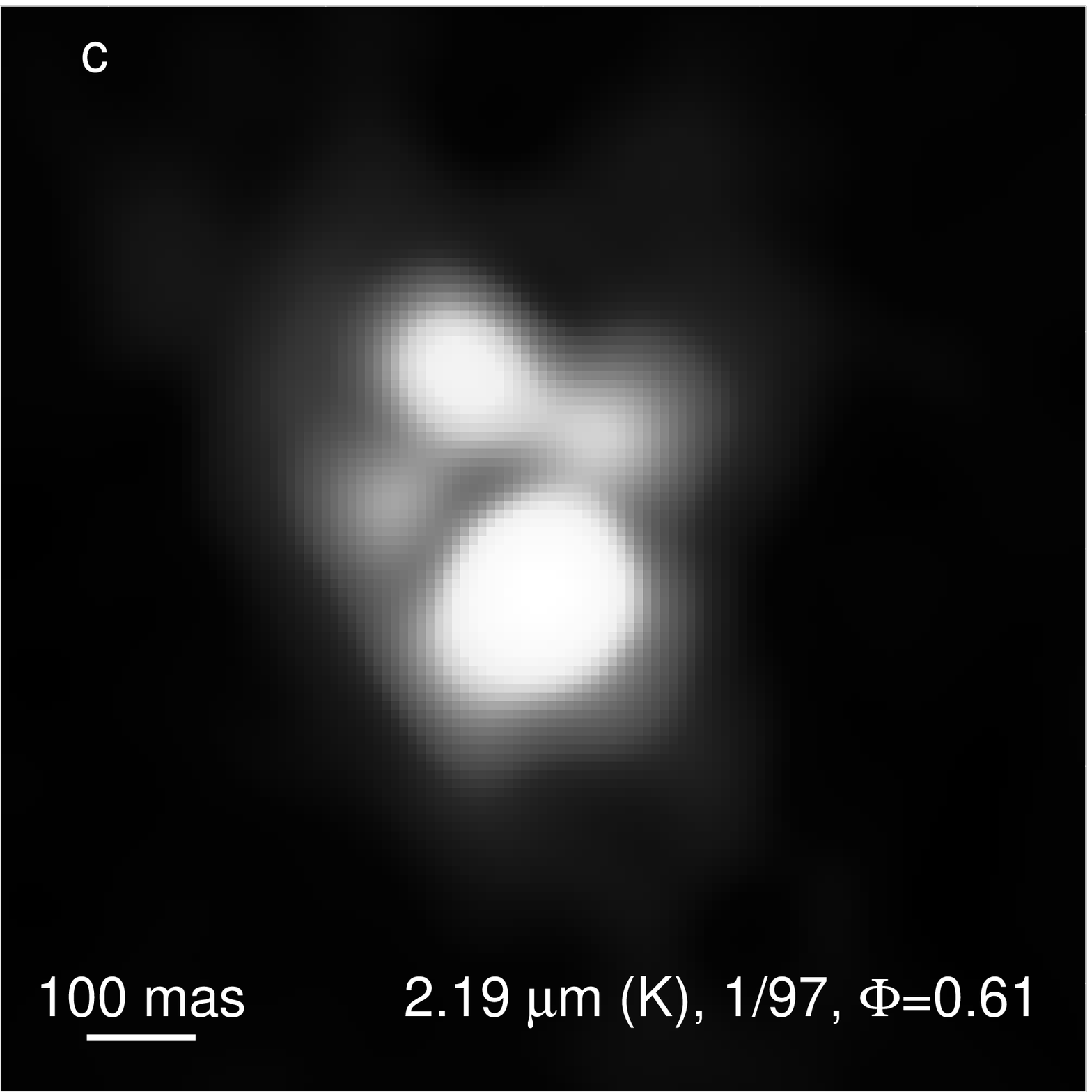}     }
    \put(0,73){  \epsfxsize=36mm\epsfbox{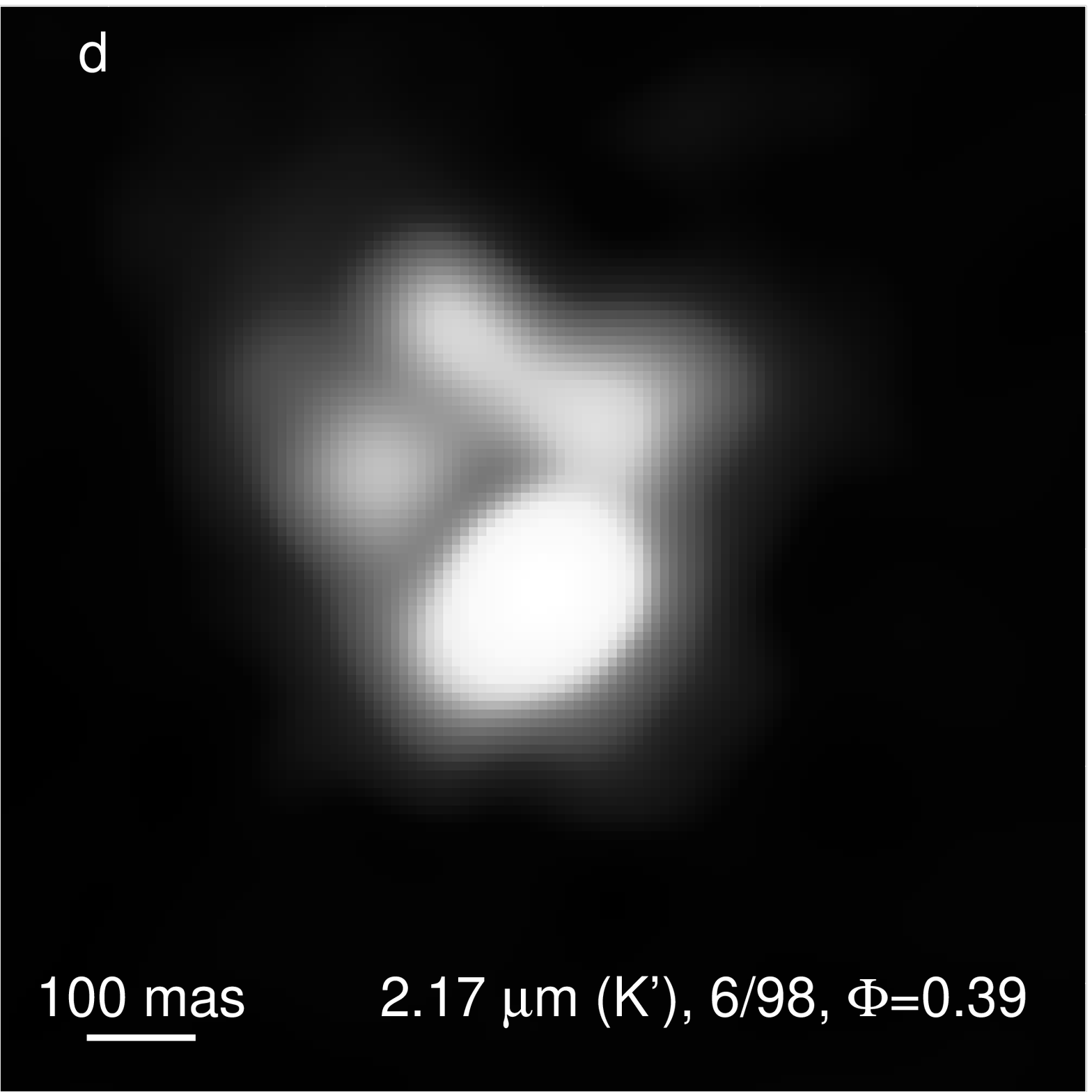}    }
    \put(0,0){   \epsfxsize=36mm\epsfbox{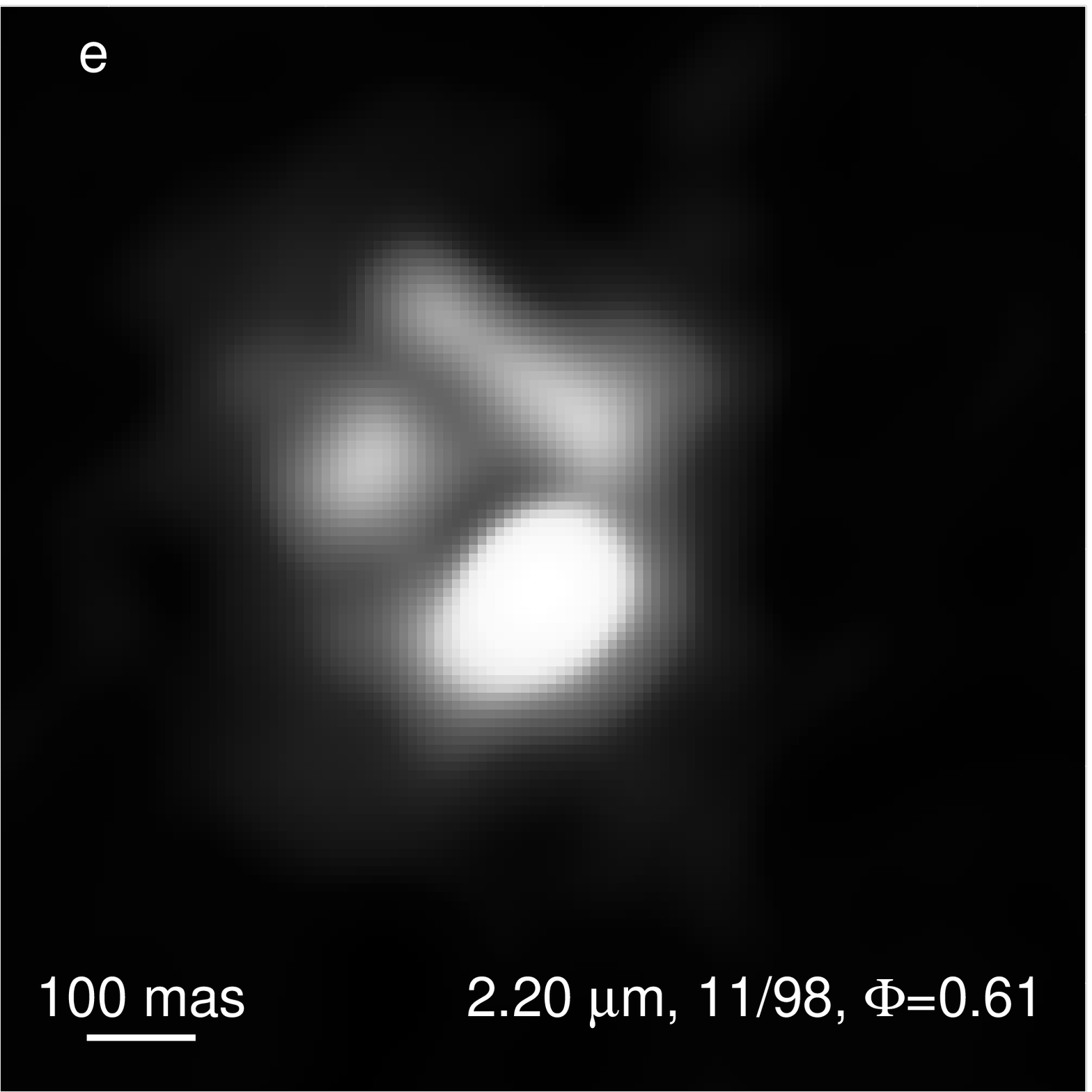}   }
    \put(73,292){ \epsfxsize=36mm\epsfbox{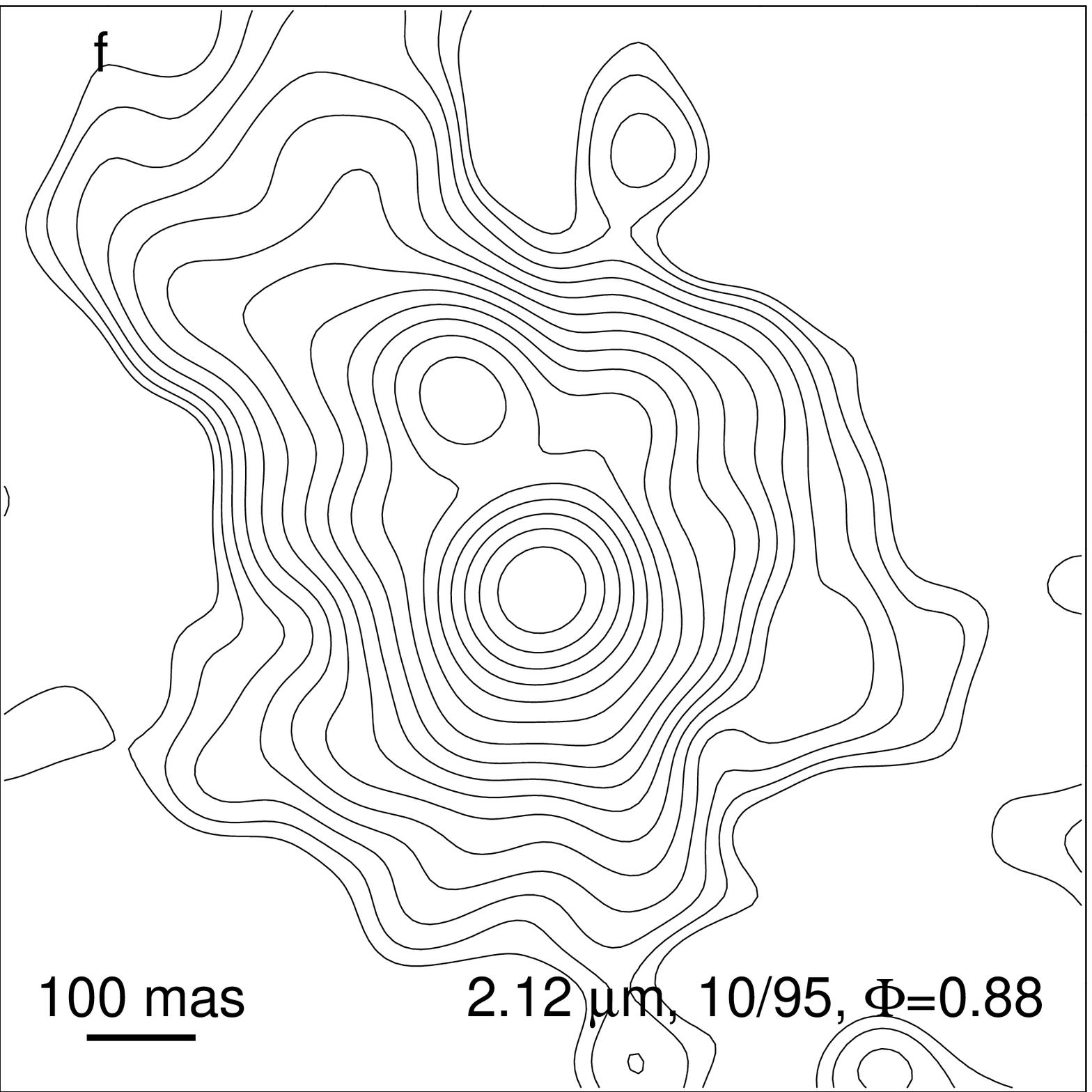}  }
    \put(73,219){ \epsfxsize=36mm\epsfbox{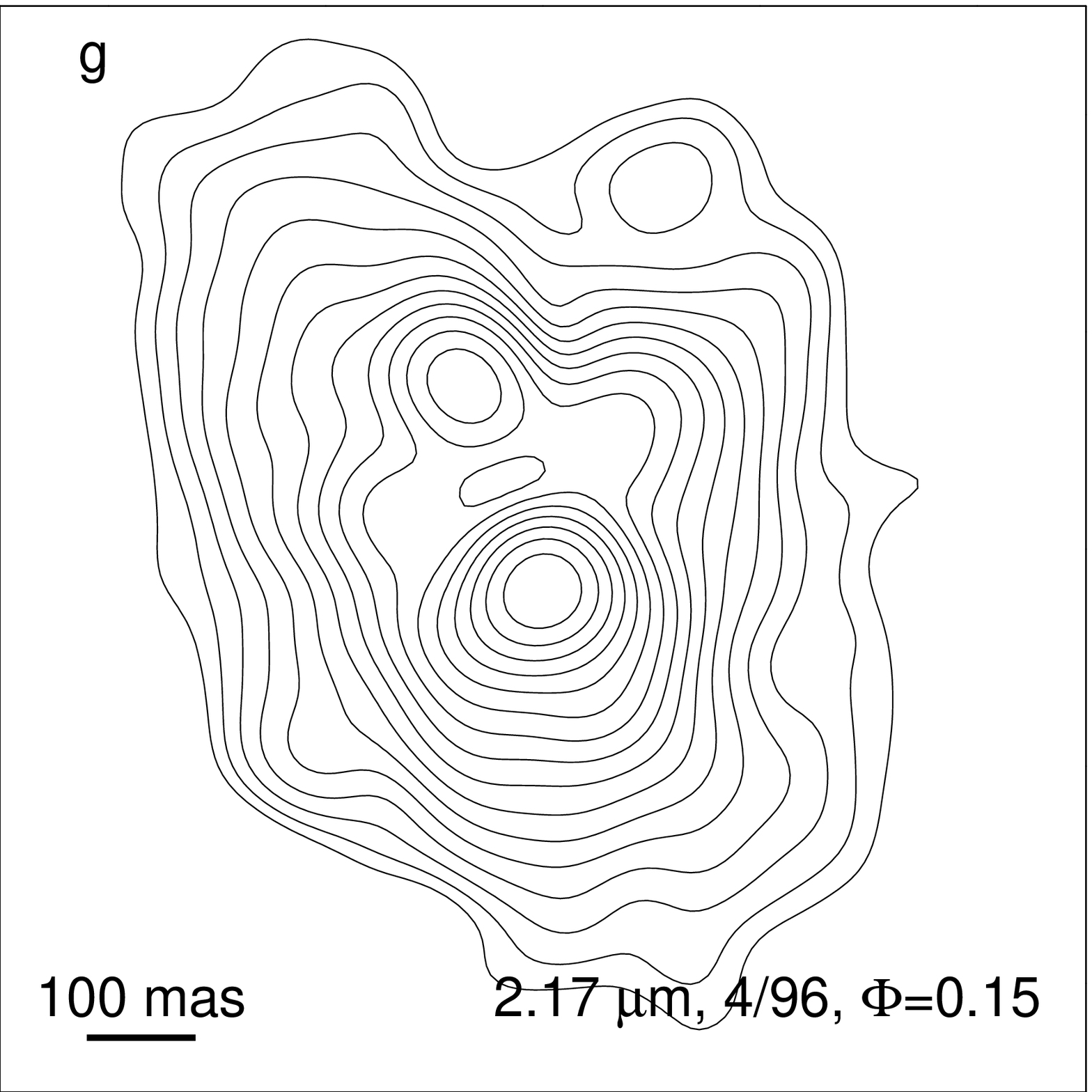}  }
    \put(73,146){ \epsfxsize=36mm\epsfbox{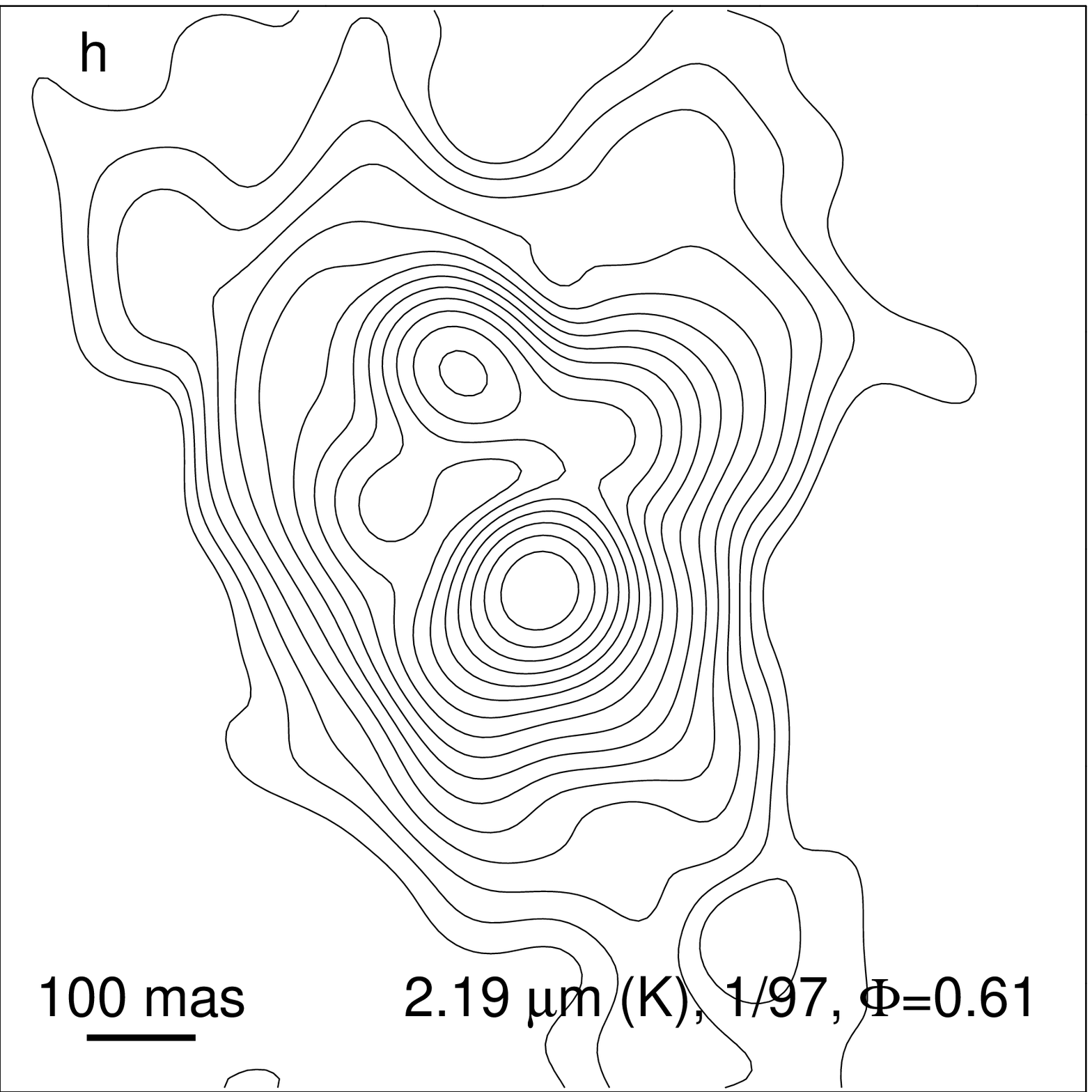}   }
    \put(73,73){  \epsfxsize=36mm\epsfbox{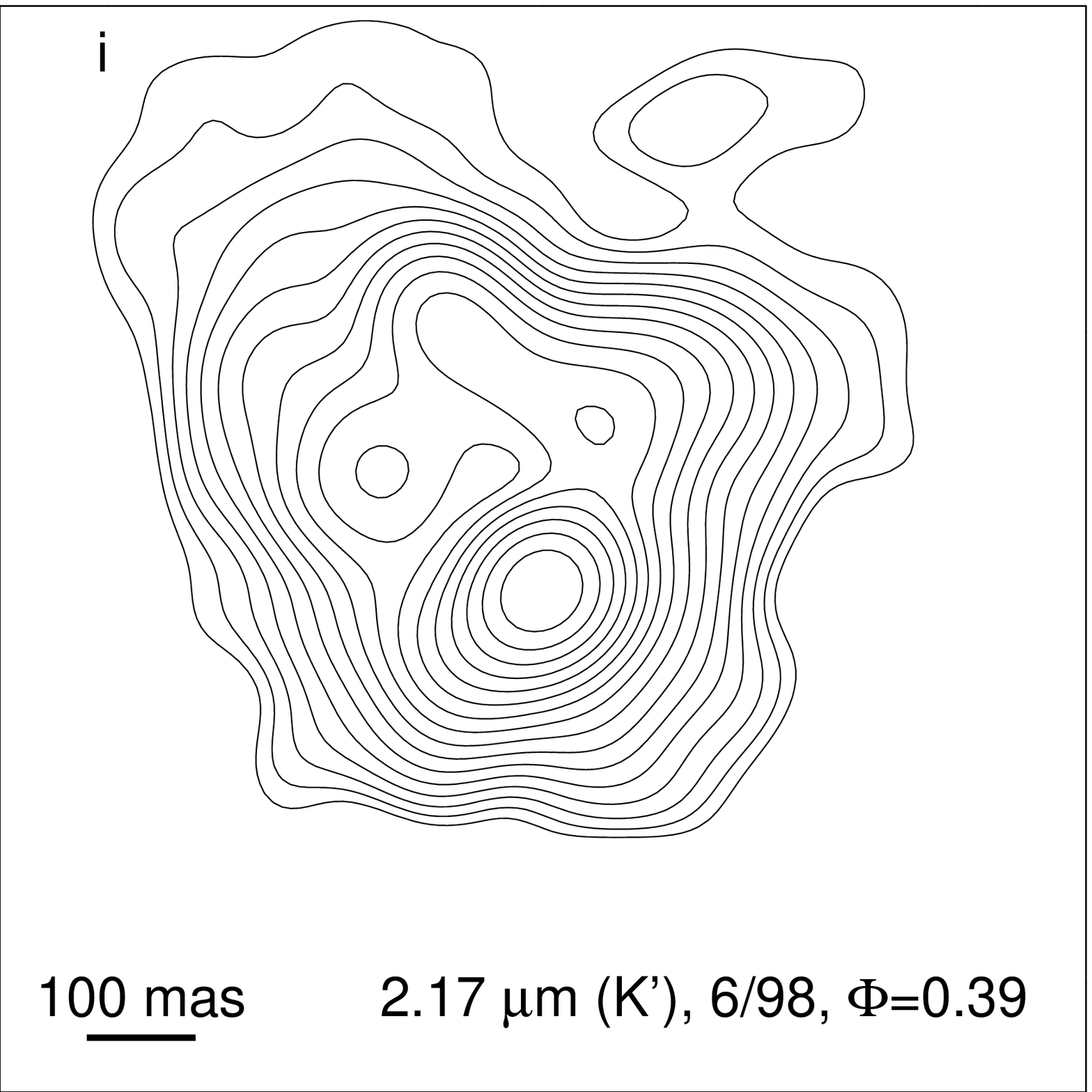}  }
    \put(73,0){   \epsfxsize=36mm\epsfbox{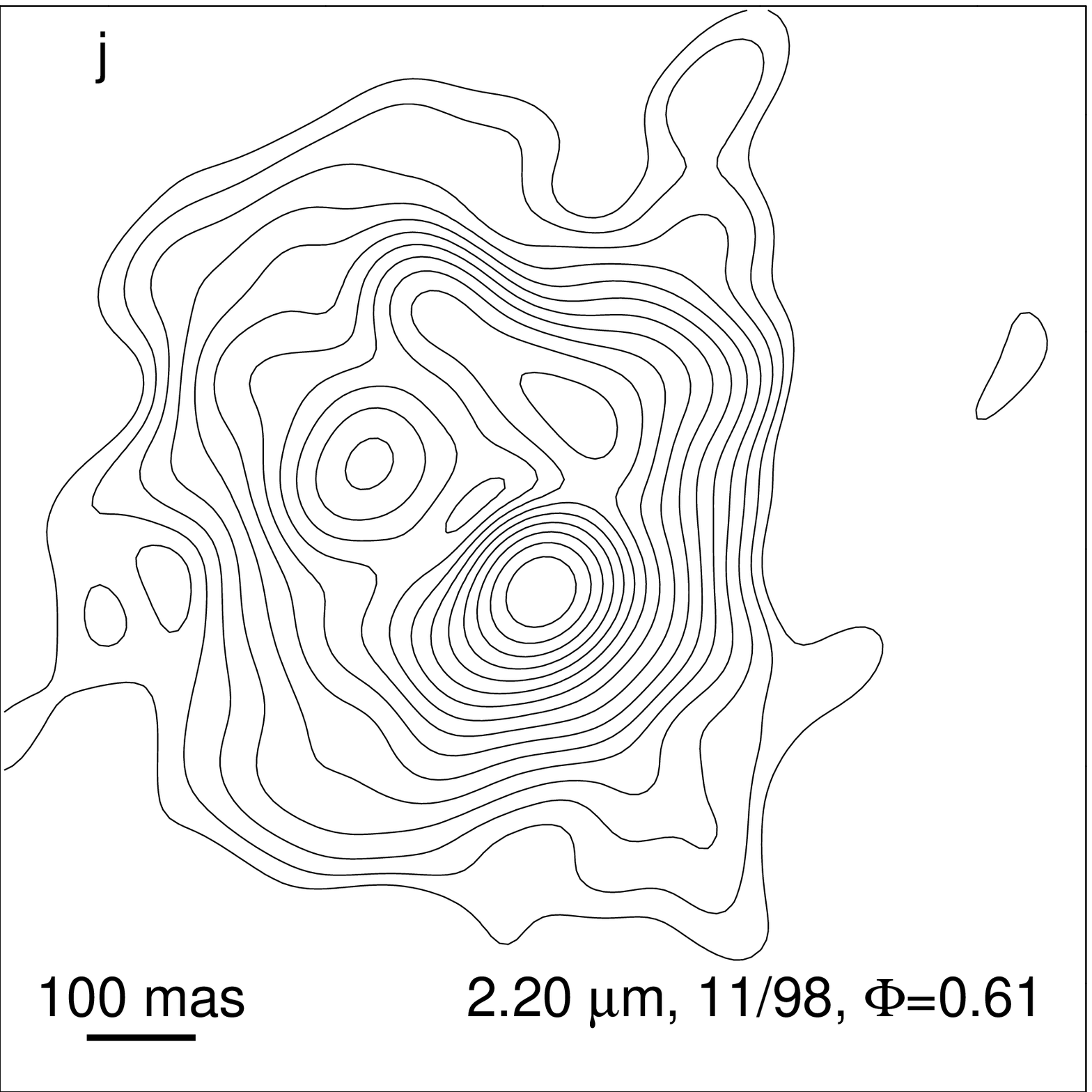} }
    \put(156,234){ \epsfxsize=55mm\epsfbox{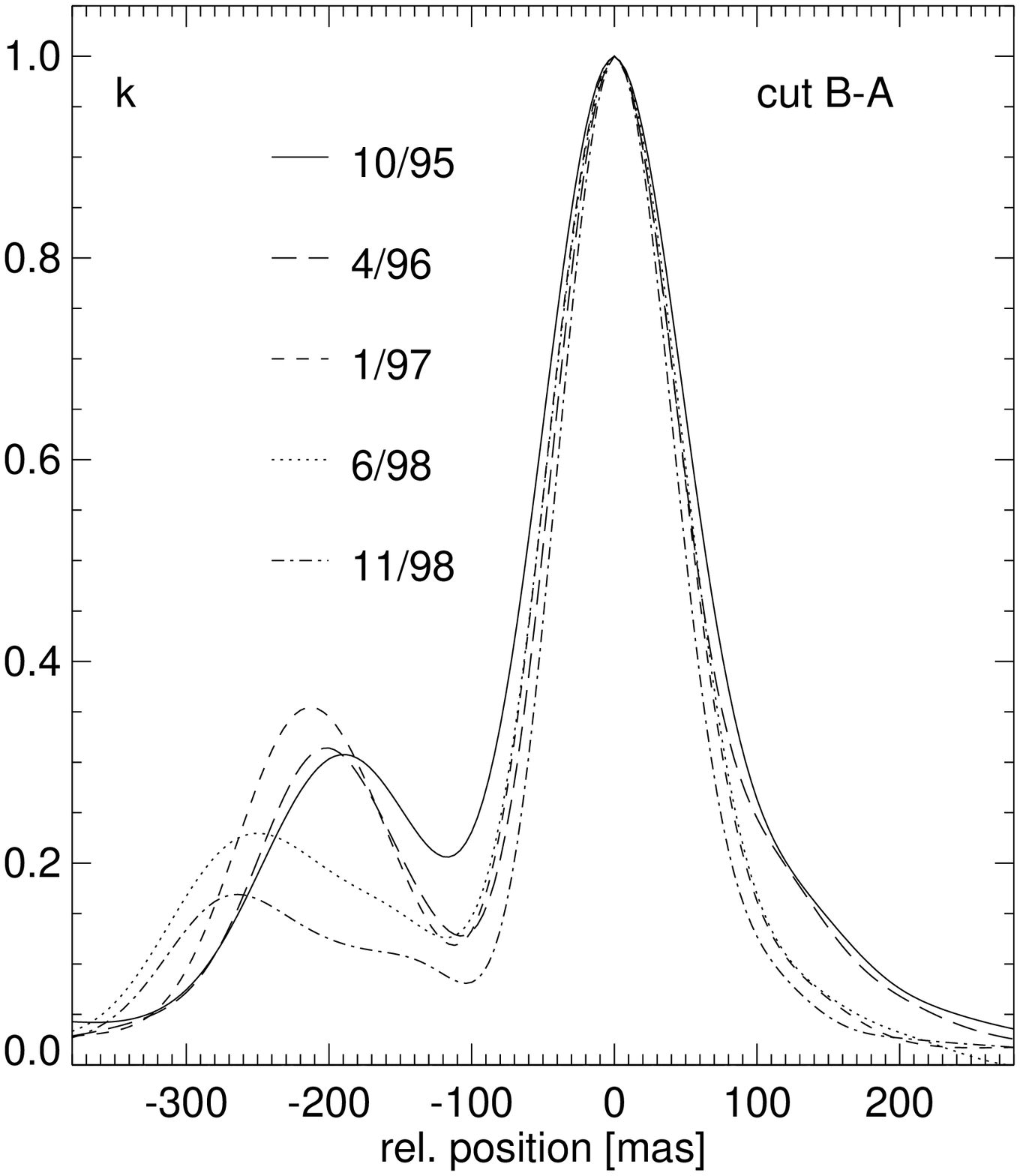} }
  \end{picture}
  \hfill
  \parbox[b]{55mm}{
  \caption{
    {\bf a} to {\bf e}:
    High--resolution bispectrum speckle interferometry images
    of IRC\,+10\,216.
    North is up and east to the left.  The figures represent a time
    series showing the evolution of the subarcsecond structure of
    IRC\,+10\,216 from 1995 (top) to 1998 (bottom).  In all
    figures the same gray level corresponds to the same relative
    intensity measured with respect to the peak.
    {\bf f} to {\bf j}:
    Same as a to e but as contour representation (contours at every 0.3~mag
    down to 4.8~mag relative to the respective peak).  The resolutions of
    the images are 92~mas (a, f), 82~mas (b, g), 87~mas (c, h), 87~mas
    (d, i), and 75~mas (e, j).
    In all figures the epoch of the observation, the filter wavelength
    and the photometric phase $\Phi$ are indicated.
    {\bf k} Cuts through the images a to e along the axis from component A to
    B (position angle 20$^\circ$).
    \label{recall}
  }
  }
\end{figure}
\begin{figure}
  \setlength{\unitlength}{1mm}
  \begin{picture}(134,44)(0,0)
    \put(0,0){  \epsfxsize=44mm\epsfbox{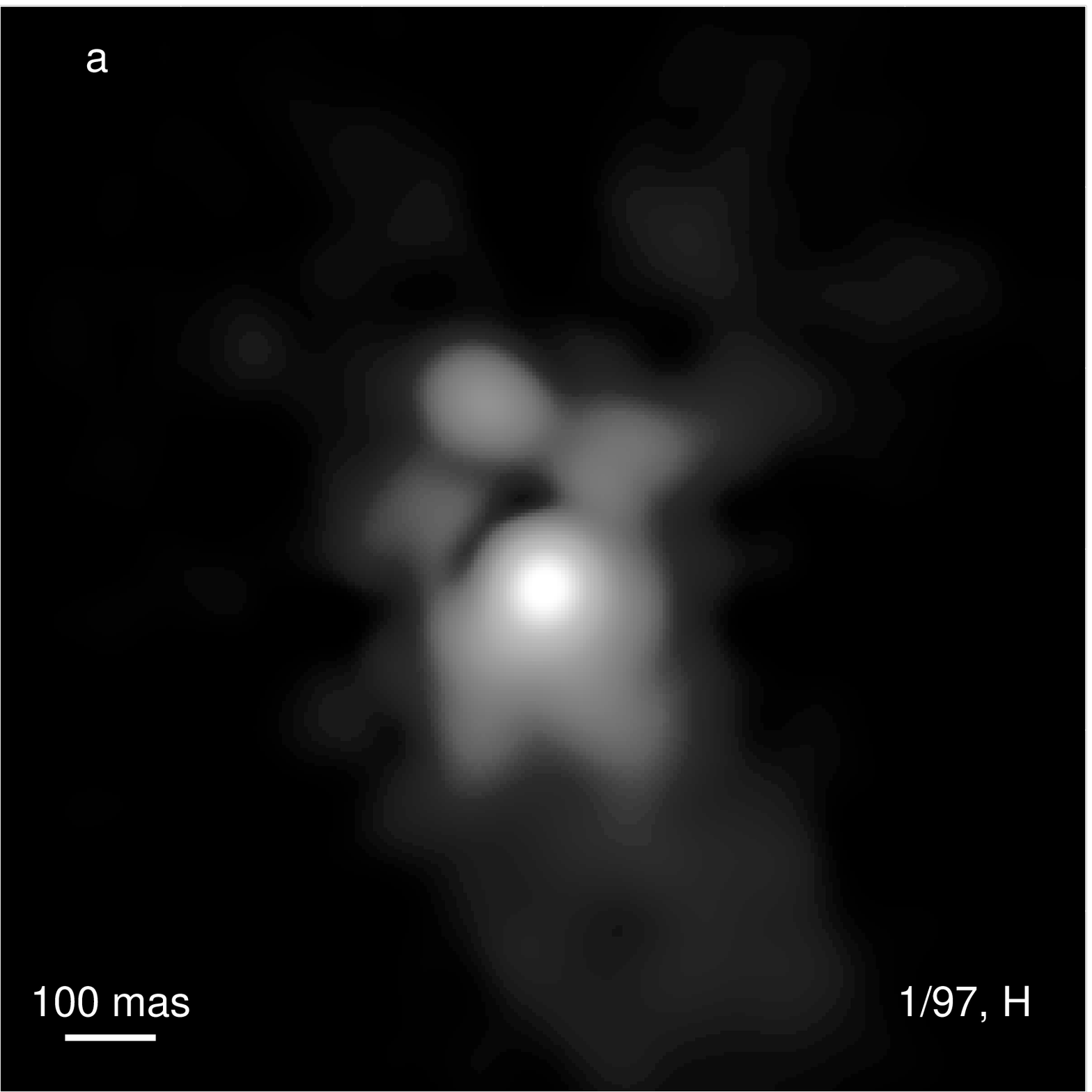}    }
    \put(45,0){ \epsfxsize=44mm\epsfbox{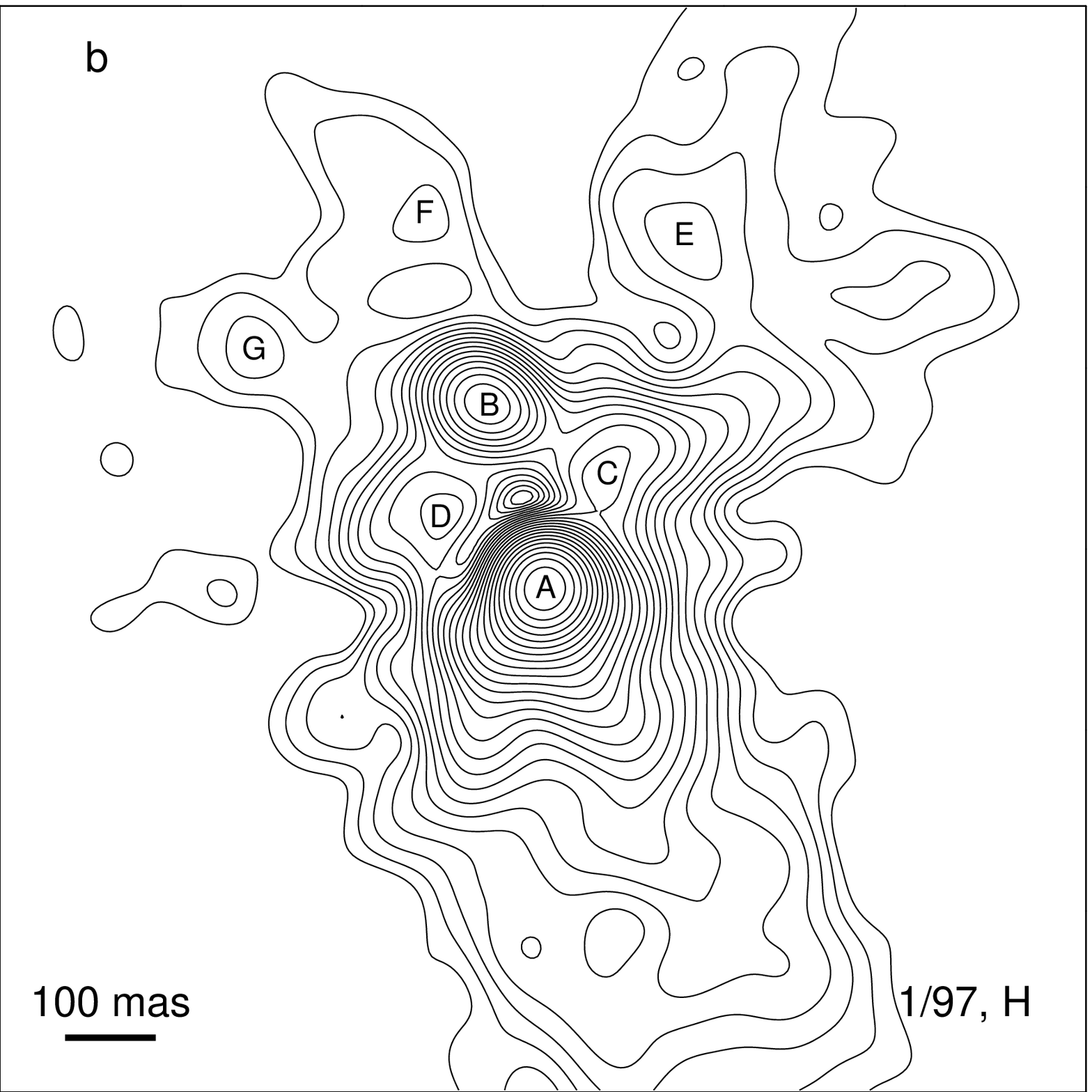} }
    \put(90,0){ \epsfxsize=44mm\epsfbox{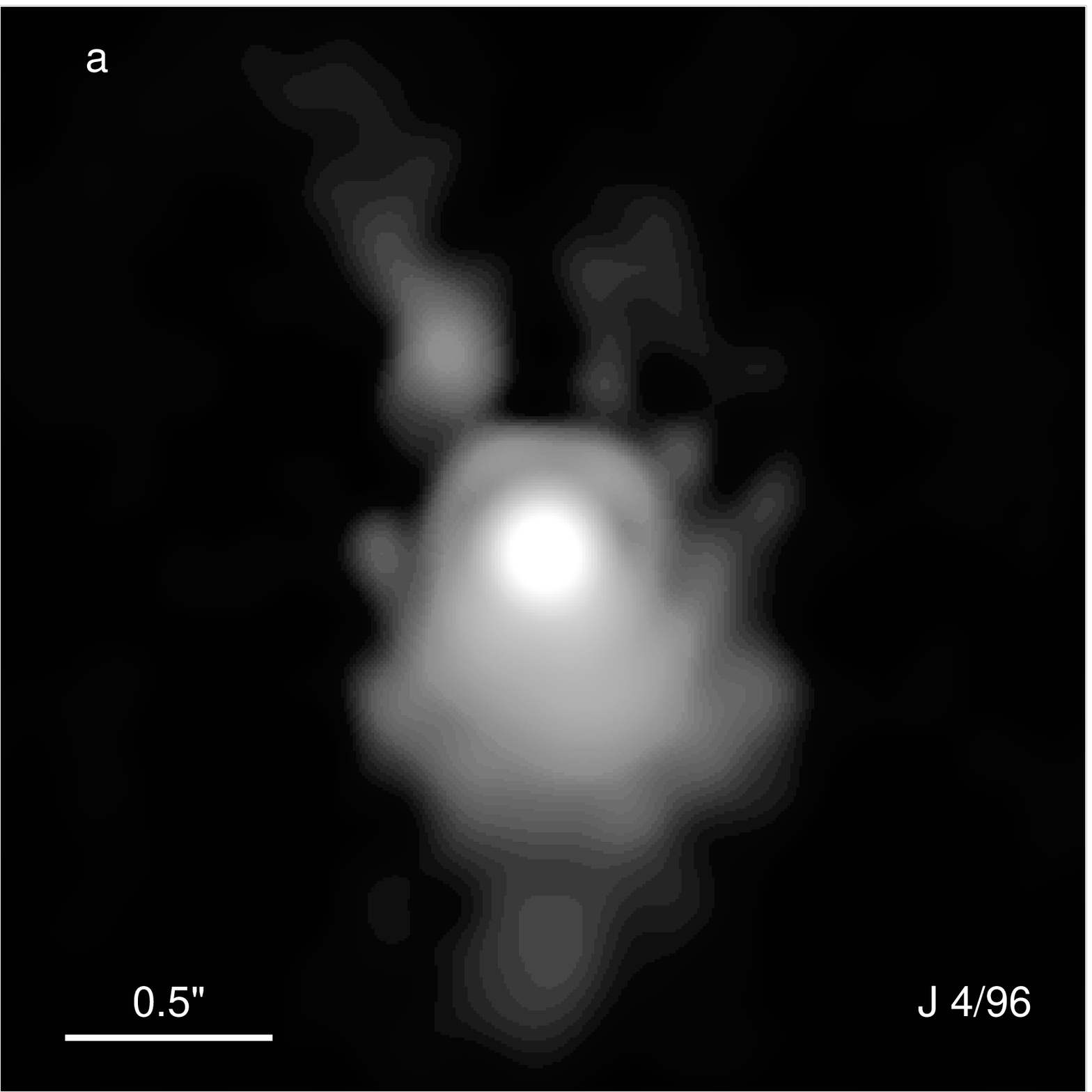}  }
  \end{picture}
  \caption{
    {\bf a} 70~mas resolution bispectrum speckle interferometry image
    of IRC\,+10\,216 in the $H$--band.  North is up and east to
    the left.
    {\bf b} Same as (a) as a contour image with denotations (A to G) of compact
    structures.  Contours are at every 0.2~mag down to 5.0~mag relative to the
    peak.
    {\bf c} $J$--band speckle reconstruction of IRC\,+10\,216 with 149~mas
	    resolution.
    \label{rech}
  }
\end{figure}
\begin{figure}
  \setlength{\unitlength}{1mm}
  \centering
  \begin{picture}(134,44)(0,0)
    \put(0,0){  \epsfxsize=44mm\epsfbox{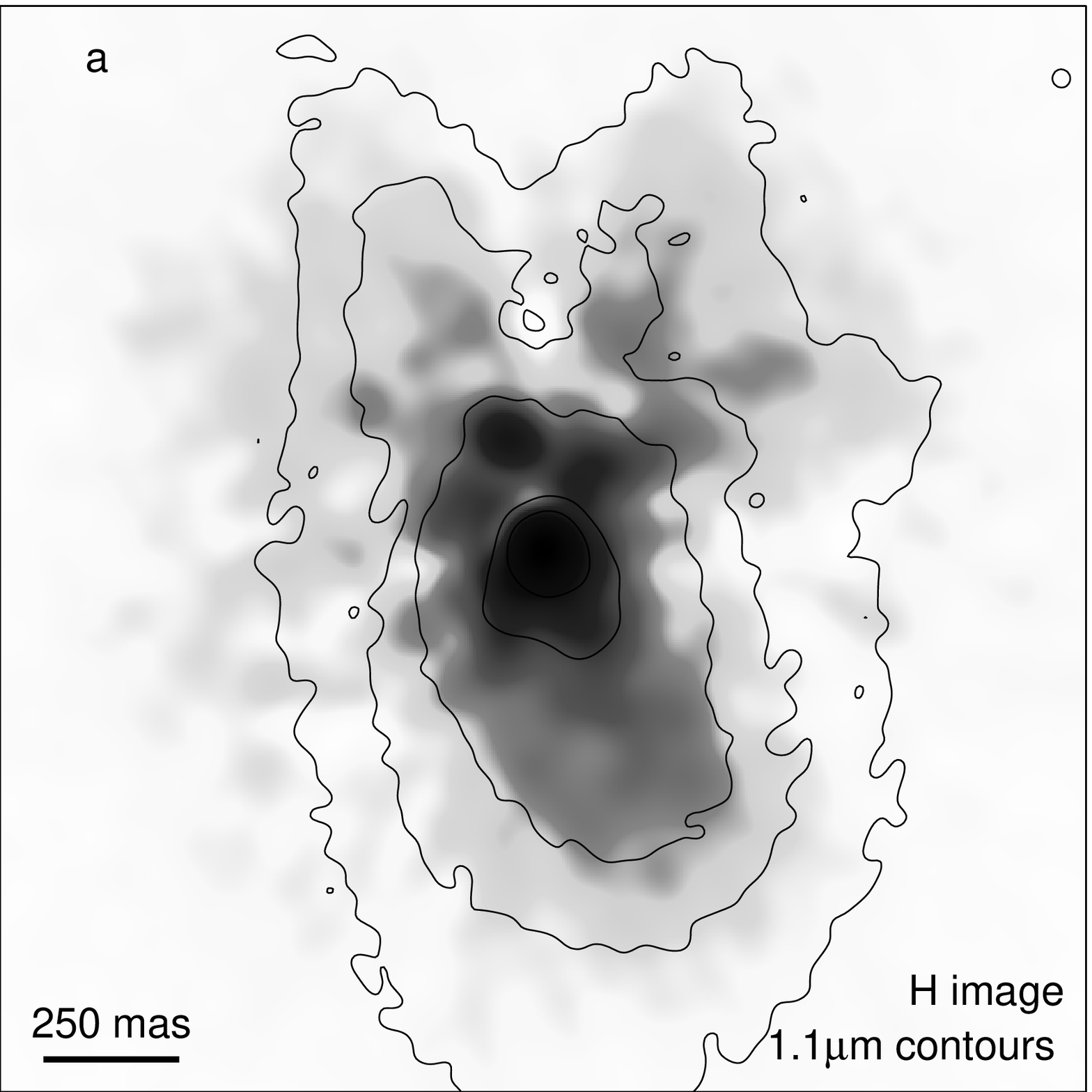}  }
    \put(45,0){ \epsfxsize=44mm\epsfbox{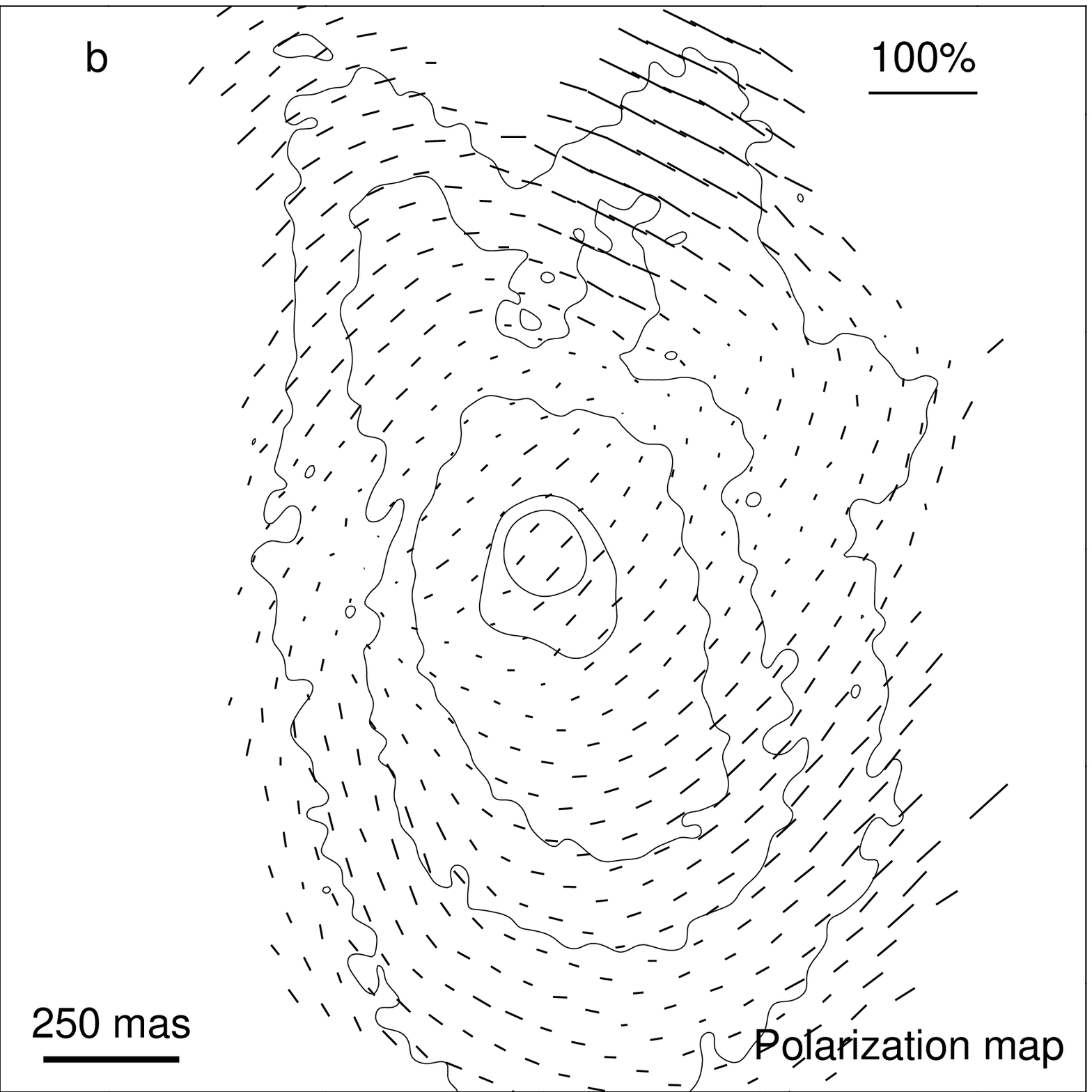}  }
    \put(90,0){ \epsfxsize=44mm\epsfbox{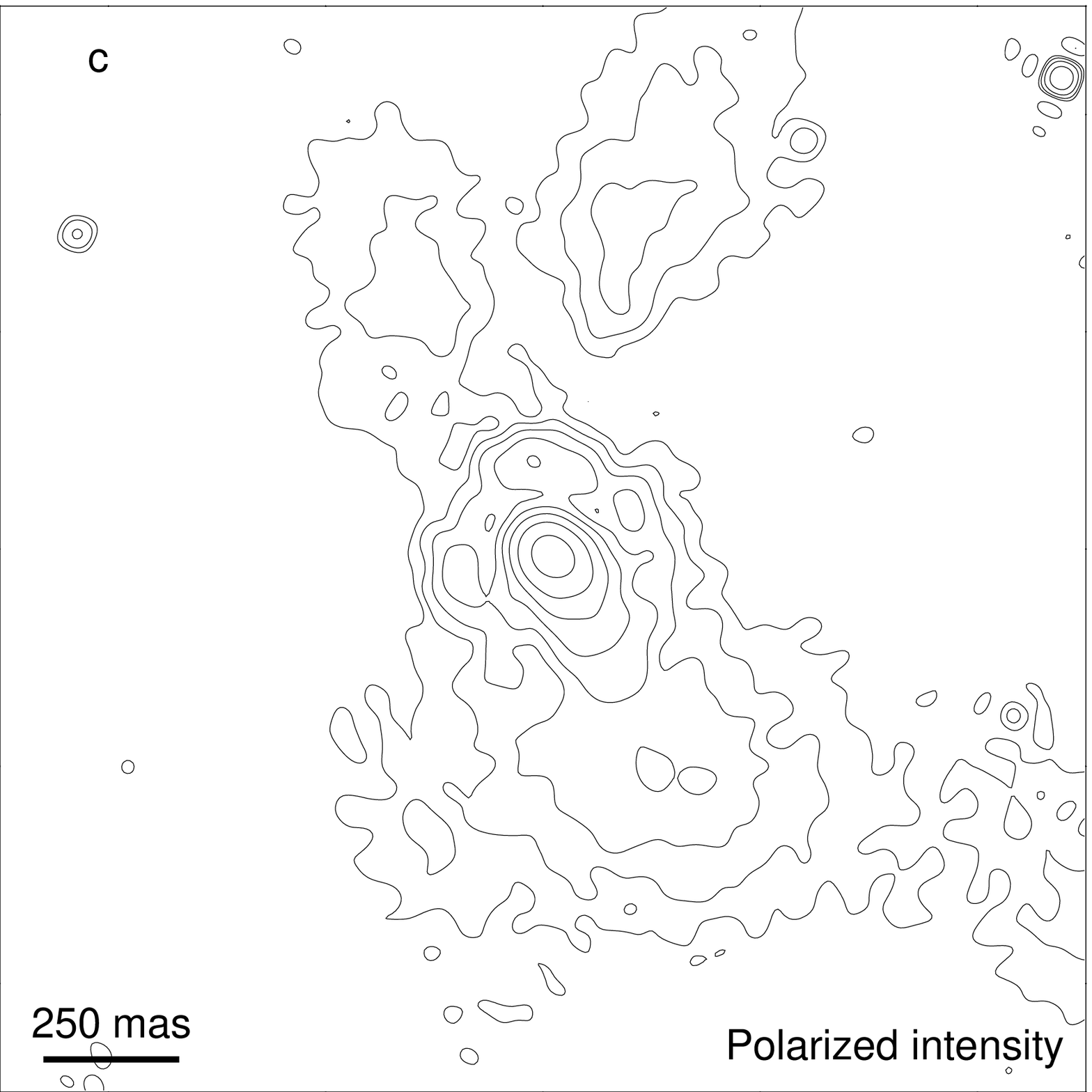}  }
  \end{picture}
  %\hfill
  \caption{
    {\bf a} Superposition of the total intensity at 1.1~$\mu$m
    (contours are at 5.0, 4.2, 3.4, 1.8, and 1.0~mag relative to
    the peak) derived from the archival HST polarization data and the
    negative $H$ speckle image.
    {\bf b} 1.1~$\mu$m polarization map.  The same contours as in
    frame {\bf a} are shown.
    {\bf c} Polarized intensity at 1.1~$\mu$m.  The contours are
    plotted from 2.4~mag to 6.0~mag difference relative to the peak of
    the total intensity in steps of 0.6~mag.
    The structure of the polarization data at the position of B is
    affected by the diffraction pattern associated with A.
    \label{polmap}
  }
\end{figure}

Figures~\ref{recall} and \ref{rech} show the $K$, $H$, and $J$ images
of the central region of IRC\,+10\,216 for all epochs.  The
high--resolution images were reconstructed from the speckle
interferograms using the bispectrum speckle interferometry method
(Weigelt 1977, Lohmann et al.\ 1983, Weigelt 1991\nocite{Weigelt91}).
We denote the
resolved components in the center of the nebula as A, B, C, and D (see
Fig.~\ref{rech}b) 
in the order of decreasing peak intensity (based on the $K$ band results
from 1996).  Figure~\ref{rech}b shows in addition three fainter
components denoted with E, F, and G.
The faint extended feature at position angle PA
$\sim$340$^\circ$ in the $J$ image corresponds quite well to the
faint component E visible in all
images in Figs.~\ref{recall} and \ref{rech} (assuming that the brightest
component in the $J$ image is coinciding with component A in the
$H$ and $K$ images, see Figure~\ref{polmap}a).

Figure~\ref{polmap} shows the results of polarimetric observations with
the HST {\sc Nicmos} camera at a wavelength of 1.1~$\mu$m (raw data
retrieved from the Hubble Data Archive, STScI).  The data were obtained
at a photometric phase of $\Phi=0.76$. 
From the data the total intensity
(Figs.~\ref{polmap}a and b), the polarized intensity
(Fig.~\ref{polmap}c), the degree and position angle of the polarization
(Fig.~\ref{polmap}b) have been derived.

\section{Nebula structures}
%\subsection{The core structure and the faint nebula in the $H$ and $K$ images}
%\label{obsneb}
%
The bright inner region of IRC\,+10\,216 is surrounded by a larger faint
nebula.  The bipolar shape of this nebula is most prominently present in
the $J$--band image (Fig.~\ref{rech}c) and in the HST images at shorter
wavelengths (Fig.~\ref{polmap}a).  However, the fact that even in the
polarized intensity the nebula is very faint on the southeastern side
suggests the main axis of the nebula to be at PA$\sim20^\circ$ to
$30^\circ$ along the direction from component A to B.  This fits well to
the main axis of the H$^{13}$CN($J=1-0$) emission (Dayal \& Bieging
1995\nocite{DayalBieging95}) which is weakly elongated on a scale of about
10$^{\prime\prime}$.
%On larger scales asymmetries are not observed (see
%e.g.\ Dayal \& Bieging 1995\nocite{DayalBieging95}, Mauron \& Huggins
%1999\nocite{MauronHuggins99}).

%
%\subsection{Separation of components A and B}
%\label{obsmotion}
%
%{\em Dynamical changes within the nebula.}
We determined the separations of the components A and B
for the 5 epochs from 1995 to 1998 shown in Fig.~\ref{recall}.
The separations are:
191~mas,
201~mas,
214~mas,
245~mas, and
265~mas.  A linear
regression fit gives a value of 23~mas/yr for the average increase in
the apparent separation of the components.  Interpreting this increase
as a real motion would lead to 14~km/s within the plane of the sky (for
a distance of $130$~pc).
The result comprises data from more than one pulsation period
($\sim$649~days, see Le~Bertre 1992\nocite{LeBertre92}) and it is thus obvious
that the apparent relative motion of the nebula components is not simply
related to the stellar variability.

Besides the motion of the components, Fig.~\ref{recall} shows
that these components change their shapes and relative fluxes.  The
brightest component A becomes narrower along the axis A--B
($\sim$20$^\circ$).  The peak--to--peak intensity ratio of B and A is
almost constant from 1995 to 1997.  Later component B is
fading.  At the same time the other components become brighter and
detached from A.  Note that the photometric phases and the
integral $K$ magnitudes of IRC\,+10\,216 in January 1997 and November
1998 are almost identical (Osterbart et~al.\
1999\nocite{OsterbartBalegaEtAl99}).  Again we find that the time scale for
the changes seen in our images is significantly different from the
period of the stellar pulsation.

\section{Discussion}
\label{discussion}
In the following we want to address the question where, behind all the
dust, the central star in the system of IRC\,+10\,216 is located.  This
question is of specific interest to understand the physical properties
of the nebula.
At short wavelengths the nebula shows a bipolar structure.  A comparison
of the observed structure to other bipolar nebulae like the Red
Rectangle (Men'shchikov et~al.\ 1998\nocite{MenshchikovBalegaEtAl98}) suggests
that the {\sf X}--like arms originate mainly from scattering of stellar
light on the surfaces of cavities.  The star then is at least partially
obscured by an optically thick dust shell or torus.  Convincingly Haniff
\& Buscher \cite{HaniffBuscher98} argued that the main axis of the
object is tilted with its southern side towards the observer.

%
%\subsection{Is component A the star?}
%\label{disa}
%
{\em Is component A the star?}
At first glance it seems reasonable to assume that the star is at the
position of the brightest component A.  However, the synthetic
polarization maps of Fischer et~al.\ \cite{FischerHenningEtAl96} show
that a significant polarization at the position of the star is only
present for nearly edge--on configurations.  The high degree of
polarization of A ($P\sim14\%$) is thus not in agreement with A being
the star because the structure of the bipolar nebula at short
wavelengths suggests that we are looking at an intermediate viewing
angle (e.g.\ 50$^\circ$ to 60$^\circ$).
At larger separations from A the
polarization pattern is centrosymmetric.  The center of such a pattern
is thought to be at the position of the illuminating source (Fischer
et~al.\ 1996\nocite{FischerHenningEtAl96}).
Since in the map in Fig.~\ref{polmap}b this center does not coincide
with A, but is located significantly north of A, we are led to the
conclusion that A is not the star.

%
%\subsection{Is the star at the position of B?}
%\label{disb}
%
{\em Is the star at the position of B?}
In the following we will show that it is consistent with the observation
to assume that the star is at the position of B.

(a) The cometary shapes of A %%% in the images in
(Fig.~\ref{rech}) as well
as the presented polarization data strongly suggest that A is part of a
scattering lobe within a bipolar structure.  Consistently, component A
and its southern tails are relatively blue ($H-K$ in the range from 2 to
3.2; see Osterbart et~al.\ 1999\nocite{OsterbartBalegaEtAl99}) compared to the
integral color ($H-K=3.2$).

(b) The northern components B, C, and D are, on the other hand, rather
red ($H-K\approx4.2$) in comparison with the integral color.  This
suggests that these structures are strongly obscured and reddened by
circumstellar dust.

(c) The brightest northern component in the $J$ image (Fig.~\ref{rech}c)
can hardly be seen in the $H$ image and is thus very blue.  This
component and A can be considered as opposite lobes of an almost
bipolar structure.  Component B is almost in the center between these
counterlobes and thus approx.\ at the position where the star would
be expected.

(d) The polarization map (Fig.~\ref{polmap}b) fits well to the picture
that the star is at B.  The center of the centrosymmetric polarization
pattern at larger separations from A is located between the two
$J$--band lobes.  This is consistent with the assumption that the
illuminating source is at or near the postion of B (cf.\ Fischer et~al.\
1996\nocite{FischerHenningEtAl96}).
Unfortunately, the polarization at the position of B itself is
contaminated by contributions from the diffraction pattern associated
with the peak A.

{\em Changes in the mass loss rate.}
The change of the shape of component A and the fading of B
can be attributed to an increasing mass loss which is
accompanied by a gradual increase of the optical depth of the dust
shell.  This is most obvious for the later observation epochs,
suggesting an enhanced mass loss since 1997.
A strongly variable mass loss has, in fact, been predicted by
theoretical models treating the dust formation mechanism in the
envelopes of long--period variable carbon stars (Winters
et~al.\ 1995\nocite{WintersFleischerEtAl95}).  Periods of this mechanism may
be significantly longer than the stellar pulsation period.
%An increasing optical depth of the inner
%dust shell also results in an increasing apparent separation of the dust
%shell structures.  The apparent motion of the components is thus not
%solely determined by the velocity of the dust particles but also by the
%changes of the optical properties of the circumstellar dust.

%
%\subsection{Alternative models: the star between A and B or near B}
%\label{disalt}
%
{\em Alternative models: the star between A and B or near B.}
It is not possible to exclude that
the star may be located between A and B, close to B, or in the center of
A, B, C, and D. The precision of the polarization map is
not sufficient to conclude whether the star is at the position of B or
only close to it. Two--dimensional radiative transfer calculations
in progess show, however, that the observed intensity ratio of A and B as well
as the components' shapes clearly require the star to be at B.
%%%%%
%{\em Radiative transfer modeling.}
%An answer to the question where the star is located
%requires radiative transfer calculations.
%Results of our two--dimensional radiative transfer modeling shows
%that the shapes of A and B cannot be reproduced when assuming a
%position of the star between A and B.  On the other hand, it was
%possible to reproduce these shapes and the intensity ratio of A and B in
%the case where the star is assumed to be at the position of B.  Clear
%preference is thus given to the latter model.

%
\section{Stellar evolution and bipolar structure}
\label{disevol}
IRC\,+10\,216 is without doubt in a very advanced stage of its AGB
evolution due to its long pulsational period, high mass-loss rate, and
carbon-rich dust-shell chemistry indicating that already a significant
number of thermal pulses did take place.
The star's initial mass can be estimated
to be $4 $M$_{\odot} \pm 1 $M$_{\odot}$ due to the observed isotopic
ratios of C, N and O in the dust shell (Guelin et~al.\
1995\nocite{GuelinForestiniEtAl95}) and the luminosity of the central star
(Weigelt et~al.\ 1998\nocite{WeigeltBalegaEtAl98}).  Accordingly, the core
mass should be $\sim 0.7$ to $ 0.8 $M$_{\odot}$ with corresponding
thermal-pulse cycle times of $\sim 1-3\,10^{4}$yr (Bl\"ocker
1995\nocite{Bloecker95}).  Introducing the mean observed mass-loss rate to
these thermal-pulse periods shows that the present stellar wind leads to
a very effective erosion of the envelope per thermal pulse cycle,
possibly as high as $\sim 1 $M$_{\odot}$/cycle.  Consequently, the whole
envelope may be lost during the next few thermal pulses leading to the
termination of the AGB evolution.
Thus, it is not unlikely to assume that IRC\,+10\,216 has entered a
phase immediately before moving off the AGB.  This is strongly supported
by the non-spherical appearence of its dust shell showing even bipolar
structures.  Unlike AGB stars, post-AGB objects as
protoplanetary nebulae often expose prominent features of
asphericities, in particular in axisymmetric geometry
(e.g.\ Olofsson 1996\nocite{Oloff96}).
Accordingly, IRC\,+10\,216 can be thought to be an object in transition.  It
is noteworthy that the establishment of bipolar structures, i.e.  the
metamorphosis into a protoplanetary nebulae, obviously already begins
during the (very end of) AGB evolution.
The clumpiness within the bipolar shape is probably due to small scale
fluctuations of the dust condensation radius which, in turn, might be
influenced by, e.g., giant surface convection cells (Schwarzschild
1975\nocite{Schwarzschild75}).  The formation of giant convection cells can be
assumed to be a common phenomenon in red giants.

The shaping of planetary nebulae can successfully be described by
interacting stellar wind models (Kwok et al.\ 1978\nocite{Kwoketal78})
Within this scenario a fast (spherical) wind
from the central star interacts with the slow wind of the preceding AGB
evolution.  The slow AGB wind is asssumed to be non-spherical
(axisymmetric) which leads to the observed morphology of planetary
nebulae (Mellema 1996\nocite{Mel96}).
%%% The cause of an aspherical AGB mass loss is still a matter of debate.
Different mechanisms to provide the required equatorial density
enhancements are discussed (cf.\ Livio 1993\nocite{Liv93}).  Among these,
binarity is one channel including common envelope evolution and spin up
of the AGB star due to the interaction with its companion
(Morris 1981\nocite{Mor81}).
Not only stellar
companions are found to be able to spin up the AGB star but also
substellar ones as brown dwarfs and planets, most effectively by
evaporation in the AGB star's envelope (Soker 1997\nocite{Soker97}).
Currently there is no observational evidence for a possible binary nature of
IRC\,+10\,216. The fact that the polarization pattern in the
southeastern part of the nebula at 1.1~$\mu$m has a different
orientation than in the rest of the nebula might be
an indication for a second illuminating source.

Mechanisms inherent to the star include rotation, non-radial pulsations,
and magnetic fields
(see e.g.\ Dorfi \& H\"ofner 1996\nocite{DorHoef96},
Soker \& Harpaz 1992\nocite{SokHar92}, Garcia-Segura et al.\ 1999\nocite{GarBer99}).
Both non-radial $p$-modes and magnetic fields appear to be only important for
significant rotation rates. Often spin-up agents due to
binarity are assumed.  For instance, Groenewegen \cite{Groenewegen96}
favours non-radial pulsation or a yet unidentified companion which spun
up the central star as the most likely explanation for the non-spherical
shape of the dust shell of IRC\,+10\,216.

AGB stars are known to be slow rotators.  Stars with initial masses
below $\sim 1.3 $M$_{\odot}$ can be expected to lose almost their
entire angular momentum during the main sequence phase due to magnetic
braking operating in their convective envelopes.  Consequently they are
not believed to develop non-spherical mass-loss due to rotation.
Stars with larger initial masses are spun down due to mass loss
but may achieve sufficiently high rotation rates at the end of AGB
evolution (Garcia-Segura et al.\ 1999\nocite{GarBer99}).  Already small
rotation rates influence dust-driven winds considerably yielding a mass
loss preferentially driven in the equatorial plane (Dorfi \& H\"ofner
1996\nocite{DorHoef96}). For supergiants leaving the Hayashi
line Heger \& Langer \cite{HegLan98} found that significant spin up of
the surface layers may take place.  Thus, at second glance, rotation might
be able to support axisymmetric mass loss during the transition to
the proto-planetary nebula phase for AGB stars as IRC\,+10\,216.

\section{Conclusions}
We have presented high-resolution $J$--, $H$--, and $K$--band observations
of IRC\,+10\,216 with the highest resolution so far at $H$ of
70~mas.  A series of $K$--band images from five epochs between October
1995 and November 1998 shows that the inner nebula is non-stationary.
The separations of the four dominant resolved components increased within
the 3 years by almost $\sim40\%$.  For the two brightest components a relative
velocity within the plane of the sky of about 23~mas/yr or 14~km/s was
found.  Within these 3 years the rather faint components C and
D become brighter whereas component B is fading.  The general geometry
of the nebula seems to be bipolar.

We find that the most promising model to explain the structures and
changes in the inner nebula is to assume that the star is at the
position of component B.  The star then is strongly but not totally obscured
at $H$ and $K$.  Consistently component B is very red in the $H-K$ color
while A and the northern $J$--band components are relatively blue. Similarly
the polarization pattern with strong polarization in the northern arms and
still a significant polarization in the peak supports this model. The
inner nebula and the apparent motions seem to be rather symmetric around
this position and the observed changes are consistent with
%%% the assumption of
an enhanced mass loss since 1997.

IRC\,+10\,216 is without doubt in a very advanced stage of its
AGB evolution. The observed bipolarity of its dust shell even reveals that
it has possibly entered the phase of transformation into a protoplanetary
nebula.

\end{document}